\documentclass[aps,preprintnumbers,superscriptaddress,showpacs,fleqn,twocolumn,longbibliography]{revtex4}
\usepackage{booktabs}
\usepackage{hyperref}
\usepackage{epsfig}
\usepackage{psfrag}
\usepackage{amsfonts}
\usepackage{amsmath}
\usepackage{graphicx}
\usepackage{float}
\usepackage{dcolumn}
\usepackage{upgreek}
\usepackage{lineno}
\usepackage{bm}

\RequirePackage[normalem]{ulem} 
\RequirePackage{color}\definecolor{RED}{rgb}{1,0,0}\definecolor{BLUE}{rgb}{0,0,1} 
\providecommand{\DIFaddbegin}{} 
\providecommand{\DIFaddend}{} 
\providecommand{\DIFdelbegin}{} 
\providecommand{\DIFdelend}{} 
\providecommand{\DIFaddbeginFL}{} 
\providecommand{\DIFaddendFL}{} 
\providecommand{\DIFdelbeginFL}{} 
\providecommand{\DIFdelendFL}{} 
\newcommand{\DIFscaledelfig}{0.5}
\RequirePackage{settobox} 
\RequirePackage{letltxmacro} 
\newsavebox{\DIFdelgraphicsbox} 
\newlength{\DIFdelgraphicswidth} 
\newlength{\DIFdelgraphicsheight} 
\LetLtxMacro{\DIFOincludegraphics}{\includegraphics} 
\newcommand{\DIFaddincludegraphics}[2][]{{\color{blue}\fbox{\DIFOincludegraphics[#1]{#2}}}} 
\newcommand{\DIFdelincludegraphics}[2][]{
\sbox{\DIFdelgraphicsbox}{\DIFOincludegraphics[#1]{#2}}
\settoboxwidth{\DIFdelgraphicswidth}{\DIFdelgraphicsbox} 
\settoboxtotalheight{\DIFdelgraphicsheight}{\DIFdelgraphicsbox} 
\scalebox{\DIFscaledelfig}{
\parbox[b]{\DIFdelgraphicswidth}{\usebox{\DIFdelgraphicsbox}\\[-\baselineskip] \rule{\DIFdelgraphicswidth}{0em}}\llap{\resizebox{\DIFdelgraphicswidth}{\DIFdelgraphicsheight}{
\setlength{\unitlength}{\DIFdelgraphicswidth}
\begin{picture}(1,1)
\thicklines\linethickness{2pt} 
{\color[rgb]{1,0,0}\put(0,0){\framebox(1,1){}}}
{\color[rgb]{1,0,0}\put(0,0){\line( 1,1){1}}}
{\color[rgb]{1,0,0}\put(0,1){\line(1,-1){1}}}
\end{picture}
}\hspace*{3pt}}} 
} 
\LetLtxMacro{\DIFOaddbegin}{\DIFaddbegin} 
\LetLtxMacro{\DIFOaddend}{\DIFaddend} 
\LetLtxMacro{\DIFOdelbegin}{\DIFdelbegin} 
\LetLtxMacro{\DIFOdelend}{\DIFdelend} 
\DeclareRobustCommand{\DIFaddbegin}{\DIFOaddbegin \let\includegraphics\DIFaddincludegraphics} 
\DeclareRobustCommand{\DIFaddend}{\DIFOaddend \let\includegraphics\DIFOincludegraphics} 
\DeclareRobustCommand{\DIFdelbegin}{\DIFOdelbegin \let\includegraphics\DIFdelincludegraphics} 
\DeclareRobustCommand{\DIFdelend}{\DIFOaddend \let\includegraphics\DIFOincludegraphics} 
\LetLtxMacro{\DIFOaddbeginFL}{\DIFaddbeginFL} 
\LetLtxMacro{\DIFOaddendFL}{\DIFaddendFL} 
\LetLtxMacro{\DIFOdelbeginFL}{\DIFdelbeginFL} 
\LetLtxMacro{\DIFOdelendFL}{\DIFdelendFL} 
\DeclareRobustCommand{\DIFaddbeginFL}{\DIFOaddbeginFL \let\includegraphics\DIFaddincludegraphics} 
\DeclareRobustCommand{\DIFaddendFL}{\DIFOaddendFL \let\includegraphics\DIFOincludegraphics} 
\DeclareRobustCommand{\DIFdelbeginFL}{\DIFOdelbeginFL \let\includegraphics\DIFdelincludegraphics} 
\DeclareRobustCommand{\DIFdelendFL}{\DIFOaddendFL \let\includegraphics\DIFOincludegraphics} 
\RequirePackage{listings} 
\RequirePackage{color} 
\lstdefinelanguage{DIFcode}{ 
  moredelim=[il][\color{red}\sout]{\%DIF\ <\ }, 
  moredelim=[il][\color{blue}\uwave]{\%DIF\ >\ } 
} 
\lstdefinestyle{DIFverbatimstyle}{ 
	language=DIFcode, 
	basicstyle=\ttfamily, 
	columns=fullflexible, 
	keepspaces=true 
} 
\lstnewenvironment{DIFverbatim}{\lstset{style=DIFverbatimstyle}}{} 
\lstnewenvironment{DIFverbatim*}{\lstset{style=DIFverbatimstyle,showspaces=true}}{} 

\begin{document}


\title{The study on the structure of exotic states $\chi_{c 1}(3872)$ via beauty-hadron decays in $pp$ collisions at $\sqrt{s}=8\,\mathrm{TeV}$}

\author{Chun-tai Wu$^{1}$, Zhi-Lei She$^1$, Xin-Ye Peng$^{1,} $\footnote{xinye.peng@cern.ch}, Xiao-Lin Kang$^1$, Hong-Ge Xu$^1$  Dai-Mei Zhou$^2$,Gang Chen$^{1} $ and Ben-Hao Sa$^3 $ }

\affiliation{ 
${}$School of Mathematics and Physics, China University of Geosciences, Wuhan 430074, China\\
${^2}$Key Laboratory of Quark and Lepton Physics, Central China Normal University, Wuhan 430079,China\\
${^3}$China Institute of Atomic Energy, P.O. Box 275(10), Beijing 102413, China}

\date{\today}
\begin{abstract}
A dynamically constrained phase-space coalescence (\footnotesize DCPC) model was introduced to study the exotic state $\chi_{c 1}(3872)$ yield for three possible structures: tetraquark state, nuclear-like state, and molecular state respectively, where the hadronic final states generated by the parton and hadron cascade model ({\footnotesize PACIAE}). The $\chi_{c 1}(3872)$/$\psi (2S)$ cross-section ratio from beauty-hadron decays (non-prompt) based on the $\chi_{c 1}(3872)$ or $\psi (2S)\to J/\psi{\pi^+}{\pi^-}$ bound state in the decay chains as a function of charged-particle multiplicity and transverse momentum in $pp$ collisions at $\sqrt{s}=8\,\mathrm{TeV}$ are calculated. A tetraquark state scenario from {\footnotesize PACIAE+DCPC} model shows better agreement with the LHCb and ATLAS measurements for the non-prompt $\chi_{c 1}(3872)$/$\psi(2S)$ cross-section ratio distributions, indicating that the $\chi_{c 1}(3872)$ is more likely to be a compact tetraquark state.

\end{abstract}

\maketitle

\section{INTRODUCTION}

In addition to mesons composed of quark-antiquark pairs and baryons consisting of three quarks, many bound states that are incompatible with traditional hadron frameworks have been observed in the decades since the quark model proposed by Gell-Mann in 1964~\cite{1}. These bound states, also known as exotic states, including multiquark states~\cite{2,3,4}, hadron molecular states~\cite{5}, hybrid states~\cite{6,7}, and glueballs~\cite{8}, are allowed and expected by the quantum chromodynamics (QCD) and the quark model. While many unconventional hadron candidates containing heavy quarks have been discovered experimentally in recent years~\cite{9}, the exact nature of even the most well-studied $\chi_{c 1}(3872)$ particle, also known as X(3872), is still unclear.

The $\chi_{c 1}(3872)$ particle as an exotic charmonium state was observed in the exclusive decay process $B^{\pm} \to K^{\pm}J/\psi\pi ^{+}\pi ^{-}$ by the Belle Collaboration in 2003, which decays into $J/\psi\pi ^{+}\pi ^{-}$~\cite{10}. Later, CDF II, D0, BESIII, and BABAR Collaboration confirmed this exotic state's discovery experimentally~\cite{11,12,13,BESIII:2013fnz}. Among them, CDF Collaboration proposed that the quantum number of $\chi_{c 1}(3872)$ particle may be $J^{PC} = 1^{++} $ or $2^{-+}$~\cite{14}, and D0 Collaboration suggested that $\psi (2S)$ state and $\chi_{c 1}(3872)$ state with the same decay mode have the same production and decay properties, which can provide a good benchmark for studying the properties of $\chi_{c 1}(3872)$ particle~\cite{12}. Finally, the spin and parity of the $\chi_{c 1}(3872)$ state are determined by the LHCb Collaboration to $J^{PC} = 1^{++} $~\cite{15}. Although there are several measurements on $\chi_{c 1}(3872)$ particle, due to the lack of the understanding of its exact properties, various models have emerged to describe $\chi_{c 1}(3872)$ state as a $D^{\ast 0}\bar{D}^{0}$ molecular state with small binding energy~\cite{16,17}, a compact tetraquark state~\cite{18,19}, a hybrid meson~\cite{20,21}, or a charmonium-molecule~\cite{22,23}.

Recently, The prompt $\chi_{c 1}(3872)$/$\psi (2S)$ cross-section ratio was measured at midrapidity by CMS Collaboration as a function of transverse momentum ($p_{\rm T}$) in Pb--Pb collisions at $\sqrt{s_{\rm NN}}=5\,\mathrm{TeV}$. The central value for the ratio is close to unity and enhanced with respect to the one measured in pp collisions~\cite{24,CMS:2013fpt}. This provides a unique experimental input to theoretical models understanding the $\chi_{c 1}(3872)$ production mechanism and the nature of its state since the modification of the hadronization mechanism is predicted when a color-deconfined state of the matter called quark–gluon plasma is formatted in heavy-ion collisions. The AMPT transport model~\cite{Zhang:2020dwn} with instantaneous coalescence, TAMU model~\cite{Wu:2020zbx} considering only the regeneration processes, the statistical hadronization model (SHM)~\cite{Andronic:2019wva} based on the assumption of thermal equilibrium predict the different magnitude of the ratio with different scenarios of the structure.

In high-multiplicity pp collisions at LHC energies, the charged-particle densities can reach values comparable with those measured in peripheral heavy-ion collisions. Measurements at the such condition in pp collisions showed features that resemble those in heavy-ion collisions~\cite{CMS:2010ifv,ATLAS:2015hzw,ALICE:2016fzo}. Recently, a multiplicity dependence of the $p_{\rm T}$-differential $\rm \Lambda_{c}^+$/$\rm D^0$ ratio is observed by ALICE Collaboration, evolving from pp to Pb--Pb collisions smoothly~\cite{ALICE:2021npz,ALICE:2021bib}.
The prompt $\chi_{c 1}(3872)$/$\psi (2S)$ cross-section ratio is found to decrease as charged-particle multiplicity increases by the LHCb Collaboration~\cite{LHCb:2020sey}, which is well described by the comover interaction model~\cite{Esposito:2020ywk}. The $\chi_{c 1}(3872)$/$\psi (2S)$ cross-section ratio from beauty-hadron decays (non-prompt) showed a slight increase trend as charged-particle multiplicity increases, no theoretical calculation is available for such measurement. Thus, studies about non-prompt $\chi_{c 1}(3872)$ production at high multiplicity $pp$ collisions can provide further insight into beauty-quark hadronization as well as an understanding of the nature of the $\chi_{c 1}(3872)$ structure.

In this paper, the $\chi_{c 1}(3872)$ from beauty-hadron decays, produced in high multiplicity $pp$ collisions at $\sqrt{s}=8\,\mathrm{TeV}$, were studied using the Monte Carlo (MC) simulation approach~\cite{29}. The multiparticle final states of $J/\psi$, $\pi ^{+}$, and $\pi ^{-}$ are generated by the parton and hadron cascade ({\footnotesize PACIAE}) model~\cite{30}. The properties of $\chi_{c 1}(3872)$ with the hadronic molecular state, the nuclear-like states, or the compact tetraquarks scenario are studied separately using the dynamically constrained phase space coalescence ({\footnotesize DCPC}) model on these bases~\cite{31,32,33,34,35,36}. With the {\footnotesize PACIAE+DCPC} model,  the non-prompt $\chi_{c 1}(3872)$ of three structures to $\psi (2S)$ cross-section ratio as a function of charged particle multiplicity and as a function of $p_{\rm T}$ were predicted.

\section{The {\footnotesize PACIAE} and {\footnotesize DCPC} model}
The {\footnotesize PACIAE}~\cite{30} model based on the {\footnotesize PYTHIA6.4}~\cite{29} is a parton and hadron cascade model that can describe multiple relativistic nuclear collisions. In this paper, the {\footnotesize PACIAE} model is used to simulate $pp$ collisions, which divides the entire collision process into four main stages: parton initiation, parton rescattering, hadronization, and hadron rescattering.

The initial-state free parton is produced by breaking the strings of quarks, antiquarks, and gluons formed in the $pp$ collision with the {\footnotesize PACIAE} model. The parton rescattering is further considered using the 2 $\to$ 2 leading-order(LO) perturbative QCD parton-parton cross sections~\cite{37}. The total and differential cross-section in the evolution of the deconfined quark matter state is calculated using the MC method. After the partonic freeze-out, the hadronization of the partonic matter is executed by the LUND string fragmentation~\cite{29} or the MC coalescence model~\cite{30}. Finally, hadron rescattering is performed based on the two-body collision until the hadronic freeze-out.

The hadron yields are calculated based on a two-step approach. Firstly, the multiplicity final states are simulated by the {\footnotesize PACIAE} model in $pp$ collisions~\cite{30}. After that, a transport model ({\footnotesize DCPC}) is introduced for the calculation of the hadron yields. The details are explained as follows.

From quantum statistical mechanics~\cite{38}, both position $\vec{q} \equiv \left ( x,y,z \right ) $ and momentum $\vec{p} \equiv \left ( p_{x},p_{y},p_{z} \right )$ of a particle cannot be defined precisely in the six-dimensional phase space,
due to the uncertainty principle,
$\mathrm{\Delta}\vec{q}\mathrm{\Delta}\vec{p}\geq h^3.$
However, a volume element $h^3$ in the six-dimensional phase space
corresponds to a state of the particle. Thus, the following integral equation can be used to estimate the yield of a single particle:
\begin{eqnarray}
Y_1=\int_{E_\alpha\le H\le E_\beta}\frac{d\vec{q}d\vec{p}}{h^3}
\label{eq: one},
\end{eqnarray}
where $E_{\alpha}$, $E_{\beta }$, and $H$ are the particle's lower and upper energy thresholds and the Hamiltonian quantity, i.e. the energy function, respectively. Furthermore, the yield of N-particle clusters or bound-state hadrons can be obtained by the following integral equation:
\begin{eqnarray}
Y_{N}=\int\cdots\int_{E_\alpha\le H\le E_\beta}\frac{d\vec{q}_{1}d\vec{p}_{1}\cdots d\vec{q}_{N}d\vec{p}_{N}}{h^{3N}}
\label{eq: two}.
\end{eqnarray}

For instance, the yield of $\chi_{c1}(3872)$ particle consisting of $J/\psi$, $\pi ^{+}$, and $\pi ^{-}$ can be calculated according to the {\footnotesize DCPC} model using the following integral formula:

\begin{eqnarray}
Y_{\chi_{c1}\left(3872\right)}=\int \dots \int {\delta_{123}\frac{d\vec{q}_{1}d\vec{p}_{1}d\vec{q}_{2}d\vec{p}_{2}d\vec{q}_{3}d\vec{p}_{3}}{h^9}}
\label{eq: three},
\end{eqnarray}
\begin{eqnarray}
\delta_{123}=\left\{\begin{array}{ll}
1  \mbox { if } 1 \equiv \pi^{+}, 2 \equiv \pi^{-}, 3 \equiv J / \psi; \\
\quad m_{0}-\Delta m \leq m_{inv }\leq m_{0}+\Delta m; \\
\quad \mathit{Max}\left\{\left | \vec{q}_{12} \right |,\left | \vec{q}_{23} \right |,\left | \vec{q}_{31} \right |\right\} \leq R_{0}; \\
0  \mbox { otherwise. }
\end{array}\right.
\label{eq: four}
\end{eqnarray}
\begin{eqnarray}
\begin{aligned}
 m_{i n v}=\left[\left(E_{1}+E_{2}+E_{3}\right)^{2} -\left(\vec{p}_{1}+\vec{p}_{2}+\vec{p}_{3}\right)^{2}\right ]^{\frac{1}{2}}
\label{eq: five}.
\end{aligned}
\end{eqnarray}

In Eq.~(\ref{eq: four}), $m_0=m_{\chi_{c 1}(3872)}=3871.69\,\mathrm{MeV/c^{2} } $ represents the rest mass of $\chi_{c 1}(3872)$ particle~\cite{39}, $R_{0}$ stands for its radius and $\Delta m$ denotes the uncertainty of the mass. $\left | \vec{q}_{12} \right |$, $\left | \vec{q}_{23} \right |$, $\left | \vec{q}_{31} \right |$ indicate the distances between each of the three component particles $\pi^{+}$, $\pi^{-}$ and $J / \psi$ under the center-of-mass system, respectively, while $\mathit{Max}\left\{\left | \vec{q}_{12} \right |,\left | \vec{q}_{23} \right |,\left | \vec{q}_{31} \right |\right\}$ represents the maximum distance taken between them. The Hamiltonian quantity $H$ satisfies the equation $H^{2}= (\vec{p}_{J/\psi}+\vec{p}_{\pi^{+}}+\vec{p}_{\pi^{-}} )^{2}+m_{inv}^{2}$, and the energy threshold upper and lower limits $E_{\alpha}$ and $E_{\beta }$ satisfy $E_{\alpha,\beta}=(\vec{p}_{J/\psi}+\vec{p}_{\pi^{+}}+\vec{p}_{\pi^{-}} )^{2}+(m_{\chi_{c 1}(3872)}\mp  \Delta m)^{2}$. Thus the dynamic constraint condition $E_\alpha\le H\le E_\beta$ in Eq.~(\ref{eq: one}) can be equivalently replaced by $m_{\chi_{c 1}(3872)}-\Delta m \leq m_{inv } \leq m_{\chi_{c 1}(3872)}+\Delta m$ in Eq.~(\ref{eq: four}).

\section{RESULTS}


The final-state hadrons, including $J/\psi$, $\pi ^{+}$, and $\pi ^{-}$, are simulated using {\footnotesize PACIAE} model in $pp$ collision at $\sqrt{s}=7$ TeV. All the parameters are fixed to the default values in {\footnotesize PACIAE} model, except parj(1), parj(2), and parj(3), which are determined by fitting data from the LHCb Collaboration for $J/\psi$, $\pi ^{+}$, and $\pi ^{-}$ in $pp$ collisions at $\sqrt{s}=7$ TeV. Here, parj(1), parj(2), and parj(3) factors are related to the suppression of diquark–antidiquark pair production compared with quark–anti-quark production, the suppression of $s$ quark pair production compared with $u$ or $d$ pair production, and the extra suppression of strange diquark production compared with the normal suppression of strange quark, respectively. With the configurations of parj$(1) = 0.10$, parj$(2) = 0.20$, parj$(3)=0.90$, the production of $J/\psi$, $\pi ^{+}$, and $\pi ^{-}$ generated by the {\footnotesize PACIAE} model fits the ALICE and LHCb data~\cite{42,43} well. Table.~\ref{yeild} summaries the comparison of the  non-prompt $J/\psi$, $\pi ^{+}$, and $\pi ^{-}$ integrated yields at the same $p_{\rm T}$ and rapidity coverage between experimental data and {\footnotesize PACIAE} model.

\begin{table}[tp]
\caption{The $J/\psi$, $\pi^{+}$ and $\pi^{-}$ yields in $pp$ collisions at $\sqrt{s}=7$ TeV calculated by the {\footnotesize PACIAE} model, compared to the ALICE and LHCb data~\cite{42,43} in $\left | y \right | < 0.5,\ 0.1< p_{T} < 3\,\mathrm{GeV/}c$ for $\pi^{\pm}$ and $2 < y < 4.5,\ 0< p_{T} < 14\,\mathrm{GeV/}c$ for non-prompt $J/\psi$, respectively.}	
\centering
\renewcommand{\arraystretch}{1.2}
\begin{tabular}{ccc} \hline  \hline
Particle &        ALICE or LHCb data~\cite{42,43}       & {\footnotesize PACIAE}\\ \hline
$J/\psi$  & $(1.60\pm0.01\pm0.23) \times 10^{-5}$ & $(1.60\pm0.03)\times10^{-5}$\\
$\pi^{+}$     & $2.26\pm0.10$                         & $2.26\pm0.01$ \\
$\pi^{-}$     & $2.23\pm0.10$                          & $2.25\pm0.03$\\ \hline \hline
\end{tabular} \label{yeild}
\end{table}

Assuming no dependence of {\footnotesize PACIAE} model paramaters between $\sqrt{s}$ = 7 and 8 TeV, the simulation was redone at 8 TeV. After that, the exotic state $\chi_{c 1}(3872)$ is constructed by the combination of $J/\psi$, $\pi ^{+}$, and $\pi ^{-}$ using {\footnotesize DCPC} model. Half of the $\chi_{c 1}(3872)$ decay width is used as the $\Delta m$ parameter, i.e,~$\Delta m=\Gamma/2=1.95\,\mathrm{MeV}$~\cite{36,44}. The $\chi_{c 1}(3872)$ can be separated into three possible structures according to $\mathit{Max}\left\{\left | \vec{q}_{12} \right |,\left | \vec{q}_{23} \right |,\left | \vec{q}_{31} \right |\right\}$. The tetraquark state, the nuclear-like state, and the molecular state are defined with the radius interval $R_{0} < 1.2\,\mathrm{fm}$ ($\chi_{c 1}^{t}$)~\cite{40}, $1.2 < R_{0} < 1.96\,\mathrm{fm}$ ($\chi_{c 1}^{n}$)~\cite{41}, and $1.96 < R_{0} < 10\,\mathrm{fm}$ ($\chi_{c 1}^{m}$), respectively.

\begin{figure}[!bp]
\includegraphics[scale=0.38]{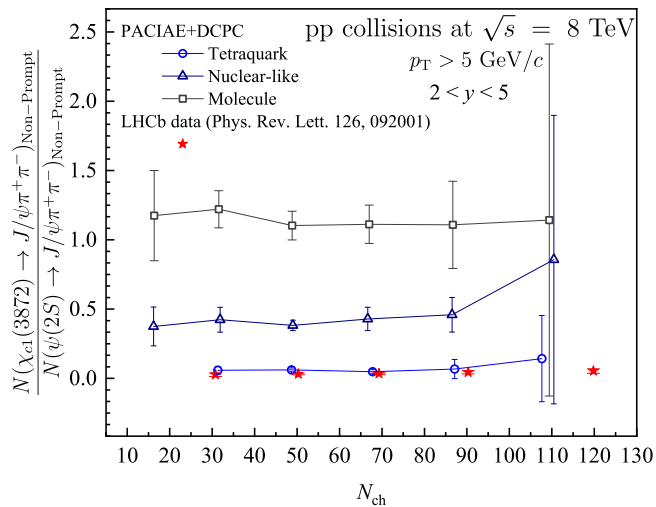}
\caption{The non-prompt $\chi_{c 1}(3872)$ to $\psi (2S)$ cross-section ratios in the $J/\psi{\pi^+}{\pi^-}$ decay channels obtained with three structures $\chi_{c 1}^{t}$, $\chi_{c 1}^{n}$, and $\chi_{c 1}^{m}$ in $pp$ collisions at $\sqrt{s}=8$ TeV, as a function of charged-particle multiplicity. The open points are computed using the {\footnotesize PACIAE+DCPC} model, and the solid red points are from the LHCb data~\cite{25}.}
\label{r-Nch}
\end{figure}

Fig.\ref{r-Nch} shows the non-prompt $\chi_{c 1}(3872)/\psi (2S)$ cross-section ratios in the $J/\psi{\pi^+}{\pi^-}$ decay channels with three structures $\chi_{c 1}^{t}$, $\chi_{c 1}^{n}$, and $\chi_{c 1}^{m}$ in $pp$ collisions at $\sqrt{s}=8\,\mathrm{TeV}$, as a function of charged-particle multiplicity ($N_{ch}$). Here, the $\psi (2S)$ yields are calculated in the same way as $\chi_{c 1}(3872)$ described above. The $N_{ch}$ represents the number of charged particles at the $2 < y < 5$ rapidity interval to match the LHCb data~\cite{25}. The non-prompt $\chi_{c 1}^{t}/\psi (2S)$ cross-section ratio is consistent with the LHCb data within uncertainties~\cite{25}, while other scenarios show larger deviation with respect to the data, indicating that $\chi_{c 1}(3872)$ is more likely to be a compact quark state. Both these three scenarios show a similar flat trend with the increasing of the $N_{ch}$ within uncertainties, consistent with the data measurement. From the {\footnotesize PACIAE+DCPC} model, the number of non-prompt $\chi_{c 1}(3872)$ naturally increases with the increasing of the multiplicity, similar to $\psi (2S)$~\cite{3}. Note that, the increasing of multiplicity will also lead to a more significant final-state effect of $\chi_{c 1}(3872)$ destruction by the comoving particles in the {\footnotesize PACIAE+DCPC} model, resulting in a decrease of the $\chi_{c 1}(3872)$ yields~\cite{26,27,28}. Similarly, $\psi(2S)$ yields are also suppressed by the final-state breakup interaction of the quarkonium with the comoving particles. However, same as argued in Ref.~\cite{25}, the effect is less pronounced for non-prompt $\chi_{c 1}(3872)$ and $\psi(2S)$ since they're produced from displaced beauty-hadron decay vertices, where the particle density is largely reduced with respect to the primary vertex.

The {\footnotesize PACIAE+DCPC} model predicts different magnitude for the non-prompt $\chi_{c 1}(3872)/\psi (2S)$ cross-section ratio based on different structures, with the hierarchy $\chi_{c 1}^{t} < \chi_{c 1}^{n} < \chi_{c 1}^{m}$. From the {\footnotesize PACIAE+DCPC} model, it's harder to generate the non-prompt $\chi_{c 1}^{t}$ with tetraquark structure with respect to other scenarios since the radius interval for the tetraquark state is smaller, it's more difficult to form the non-prompt $\chi_{c 1}^{t}$ in the limited phase space via coalescence mechanism.

\begin{figure}[htbp]
\includegraphics[scale=0.38]{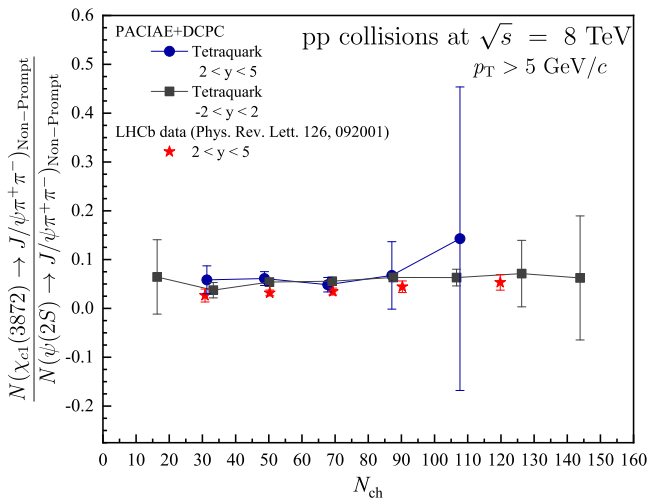}
\caption{The comparison of the non-prompt $\chi_{c 1}(3872)/\psi (2S)$ cross-section ratio as a function of charged-particle multiplicity at middle ($-2 < y < 2$) and forward rapidity ($2 < y < 5$) from the {\footnotesize PACIAE+DCPC} model, compared to the LHCb data at forward rapidity~\cite{25}. The blue and black points represent the {\footnotesize PACIAE+DCPC} model results in middle and forward rapidity, respectively, and the solid red points are from the LHCb data at forward rapidity~\cite{25}.}
\label{r-y}
\end{figure}

A natural next step would be to study the properties of $\chi_{c 1}(3872)$ as a compact tetraquark state, as the rapidity and $p_{\rm T}$ dependence of the non-prompt $\chi_{c 1}(3872)/\psi (2S)$ cross-section ratio may give further insight into beauty-quark hadronization. Fig.~\ref{r-y} reports the non-prompt $\chi_{c 1}(3872)/\psi (2S)$ cross-section ratio with tetraquark scenario as a function of charged-particle multiplicity at middle rapidity ($-2 < y < 2$) and forward rapidity ($2 < y < 5$), compared to the LHCb data at forward rapidity~\cite{25}. The results from the {\footnotesize PACIAE+DCPC} model indicates minor rapidity dependence for the non-prompt $\chi_{c 1}(3872)/\psi (2S)$ cross-section ratio.
\begin{figure}[htbp]
\includegraphics[width=0.45\textwidth]{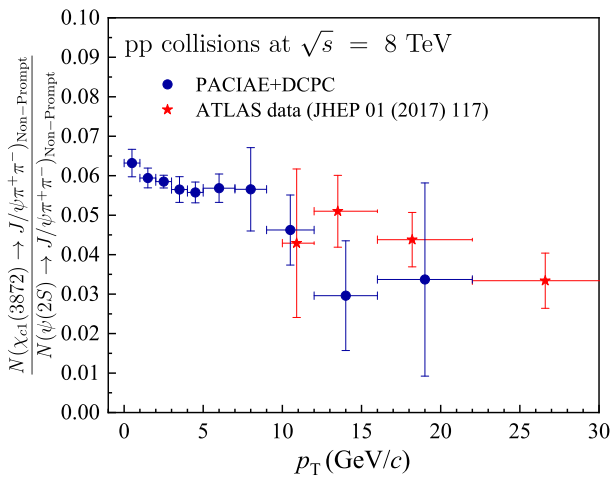}
\caption{The non-prompt $\chi_{c 1}(3872)/\psi (2S)$ cross-section ratio as a function of $p_{\rm T}$ in $pp$ collisions obtained with the {\footnotesize PACIAE+DCPC} model at $\sqrt{s}=8\,\mathrm{TeV}$, compared with the ATLAS data~\cite{24}.}
\label{R_Pt}
\end{figure}

The non-prompt $\chi_{c 1}(3872)/\psi (2S)$ cross-section ratio with tetraquark scenario as a function of $p_{\rm T}$ at middle rapidity is presented in Fig.~\ref{R_Pt}. The result is compared with the ATLAS measurement~\cite{24}. In the common interval $10 < p_{\rm T} < 22$ GeV/$c$, the result from the {\footnotesize PACIAE+DCPC} model shows a good agreement with the ATLAS data within uncertainties. The model result predicts a slightly increasing trend toward low $p_{\rm T}$, mainly due to the larger coalescence probability for $\chi_{c 1}(3872)$ at low $p_{\rm T}$ region. Nevertheless, the decay kinematic effect may also play a role due to the mass difference between the parent-beauty hadron and non-prompt hadron for these two particles~\cite{ALICE:2022tji}, which is hard to isolate for such non-prompt hadron measurements.

\section{conclusions}

In summary, the {\footnotesize PACIAE} model is used to generate final-state particles in $pp$ collisions at $\sqrt{s}=8\,\mathrm{TeV}$. The $\pi^{+}$, $\pi^{-}$, and $J/\psi$ originating from beauty-hadron decays are inserted into the {\footnotesize DCPC} model to produce the exotic state $\chi_{c 1}(3872)$. With different spatial parameters $R_0$ selected, the exotic states $\chi_{c 1}(3872)$ of three different structures are constructed as compact tetraquark state, nuclear-like state, and molecular state, respectively. The non-prompt $\chi_{c 1}(3872)/\psi (2S)$ cross-section ratios in the $J/\psi{\pi^+}{\pi^-}$ decay channels with the three structures as a function of charged-particle multiplicity are obtained from the {\footnotesize PACIAE+DCPC} model, the compact tetraquark state scenario describes the LHCb data well, indicating that the $\chi_{c 1}(3872)$ is more likely to be a compact quark state. Meanwhile, the {\footnotesize PACIAE+DCPC} model predicts a minor rapidity dependence and a decreasing trend with the increasing of the $p_{\rm T}$ for the ratio, indicating that the coalescence mechanism may play an important role in the beauty-quark hadronization in a small system. In particular, the slightly decreasing trend with the increasing of the $p_{\rm T}$ for the non-prompt $\chi_{c 1}(3872)/\psi (2S)$ cross-section ratio predicted by {\footnotesize PACIAE+DCPC} model at low $p_{\rm T}$, can be further tested with the ongoing high luminosity Run 3 data at the LHC by multi-experiments.

\textbf{Acknowledgments}
The work of X. Y. Peng is supported by the NSFC Key Grant 12061141008 and the National key research and development program of China under 2018YFE0104800. the National Natural Science Foundation of China (NSFC) (Grant No. 12205259), and of X. L. Kang is supported by the NSFC (12005195).

\nocite{*}

\bibliography{apssamp}

\end{document}